# A Relaying Incentive Scheme in Multihop Cellular Networks Based on Coalitional Game with Externalities

Cuilian Li[1,2], Zhen Yang[1], Feng Tian[1]

(1.Institute of Signal Processing and Transmission, Nanjing University of Posts and Telecommunications, Nanjing, Jiangsu 210003, China; 2. Zhejiang Wanli University, Ningbo, Zhejiang 315100, China)

***Abstract***[1]—Cooperative multihop communication can greatly increase network throughput, yet packet forwarding for other nodes involves opportunity and energy cost for relays. Thus one of the pre-requisite problems in the successful implementation of multihop transmission is how to foster cooperation among selfish nodes. Existing researches mainly adopt monetary stimulating. In this manuscript, we propose instead a simple and self-enforcing forwarding incentive scheme free of indirect monetary remunerating for asymmetric (uplink multihop, downlink single-hop) cellar network based on coalitional game theory, which comprises double compensation, namely, ***Inter-BEA***, global stimulating policy allotting resources among relaying coalitions according to group size, and ***Intra-BEA***, local compensating and allocating rule within coalitions. Firstly, given the global allotting policy, we introduce a fair allocation estimating approach which includes remunerating for relaying cost using Myerson value for partition function game, to enlighten the design of local allocating rules. Secondly, given the inter- and intra-BEA relay fostering approach, we check stability of coalition structures in terms of internal and external stability as well as inductive core. Theoretic analysis and numerical simulation show that our measure can provide communication opportunities for outer ring nodes and enlarge system coverage, while at the same time provide enough motivation with respect to resource allocation and energy saving for nodes in inner and middle ring to relay for own profits.

***Index Terms***—*Multihop cellular networks, Forwarding incentive, Partition function game, Restricted cooperation, Myerson value*

## 1  Introduction

Multihop cellular network (MCN [1, 3, 4]) is considered as a promising candidate of fourth generation (4G) wireless network for future mobile communications.

[1] Supported by National Natural Science Foundation of China (No.60772062), the Key Projects for Science and Technology of MOE (No.206055) and the Key Basic Research Projects for the Natural Science of Jiangsu Colleges (No.06KJA51001).

The fundamental idea of multihop communication is to break an original long communication link into two or more shorter links, and thus could reduce the required transmission power of each node participating in the communication scenario. The reduced transmission power could also lead to a lower interference level and shorter frequency reuse distance. In addition, the need for short-range transmission in MCNs opens the possibility of using other higher data rate wireless technologies such as IEEE 802.11, Bluetooth, or Ultra-Wideband (UWB), in conjunction with the cellular technology.

In a MCN, data packets may have to be relayed hop by hop from a given mobile station (MS) to a base station (BS) [2]. The sacrifice of a relay is two-dimensional, namely the opportunity cost [8], associated with sharing own bandwidth, and energy loss. In view of this, MSs apparently need tangible incentives to cooperate. Most existing researches use monetary pricing and economic incentives to foster packet forwarding, though the monetary reward may come in different forms: as credit counters in [5], as billing accounts in [6], as micro-payment (i.e., electronic tokens) in [7], as a 'top-up' of the credit value of a smart card in [9], or as a discount in the user subscription price for accessing the network [10]. By way of exception, M. Lindstrom etc. proposed to stimulate forwarding by pricing and rewarding with the addition of allowing an originating source to delegate part of its bandwidth to a relay for forwarding its traffic in [8]; Wei Hung-Yu etc. in [26] propose a scheduling algorithm as an incentive mechanism for hybrid wireless relay network based on Nash Equilibrium [28]. In addition to a central operator that maintains a billing account for each node, these economic rewarding methods generally require particular records to be maintained and manipulated in each node, thus involve certain security and credit problem. Furthermore, due to their indirectness and lag in reward, these measures are not suitable in changeful scenes such as emergency and temporary cases with frequent entry and quitting.

The objective of this paper is to design a simple and direct mechanism free of indirect monetary rewards to encourage cooperative relay in a MCN system. Our approach makes double compensation for potential cost of relays in opportunity and energy, i.e., stimulating policies of BS and further rewards from source nodes. Firstly, BS announces a resource (e.g., channel) apportioning policy that would allot far more share to relay-source groups than to individual nodes, thus makes it profitable for MSs to form relaying groups; secondly, stimulated by the above policy, MSs seek to form cooperative groups and source(s)-relay(s) within a group can share extra cooperative payoffs (allotted channels) by bargaining, or in some fair or stable criteria. In our implementation, multihopping (MH) is self-enforced through stimulating policy from the operator and selfish and rational reaction from the involved MSs.

The remainder of this paper is organized as follows. Section 2 describes system configuration and the global inter-coalition stimulating policy according to group size: Inter-BEA. Section 3 introduces our relaying cooperative model and analyses its properties. Section 4 estimates possible fair allocation value based on compensated Myerson value and proposes a feasible local resource allotting rule within coalitions given BS policy: Intra-BEA. We then discuss in Section 5 which cooperation structure is preferred under given BEA allocation approach. We propose coalition formation algorithm and numerical results in Section 6, so as to examine effectiveness of the proposed incentive measure. Section 7 concludes the paper.

## 2 System configuration & global policy

### 2.1 Asymmetrical MCN based on OFDMA

As uplink coverage in a cellular system is limited by power of the terminals, cellular coverage is typically asymmetrical in up- and downlink. Consider an asymmetrical MCN (single-hop down-link, multi-hop up-link) shown as Fig.1, suppose that the infrastructure is capable of providing universal coverage for area (radii of the area denoted as $r_{BS}$) of all three rings, while uplink coverage is the inner two rings (only MSs in this area can reach BS with a single-hop direct connection, though they may choose multihop for the sake of saving energy). MSs in outer ring can only reach BS by multihop through MSs in inner or middle ring.

Assume that the MCN system is orthogonal frequency division multiple access (OFDMA) based. To foster relaying cooperation and thereby improve coverage, BS announces an apportioning policy that allotting sub-channels according to cooperative group size (suppose in sheer single-hop (SH) mode, BS apportions all available sub-channels among all reachable active MSs (i.e., those within inner and middle ring, to equalize their share). In MCN mode, apart from SH transmitting, the system also allows following uplink modes: 3-hop transmitting (e.g., CIV in Fig.1, i.e., relaying cooperation among 3 MSs in inner, middle and outer ring respectively) and 2-hop transmitting, including type CI (cooperation between MSs in inner and middle ring respectively), type CII (between MSs in middle and outer ring) and CIII (between MSs in inner and outer ring).

## 2.2 Proposed Inter-coalition incentive strategy: Inter-BEA

Generally speaking, a global resource policy aims at encouraging cooperation should meet two metrics, namely efficient (i.e., it should be able to make the most of all available resource) and effective (i.e., it should be able to provide cooperating motivation for nodes in the system). Intuitively, nodes will prefer to stay (or join in) a group as long as $\lambda_n > n\lambda_1$ (i.e., a node can averagely get more in a group than staying as a singleton), namely, such an apportioning policy is effective. Denote the amount of resources allotted to a group of size n (i.e., it has n members) as $\lambda_n$, we propose an incentive measure as follows: Normalize the total resources (number of sub-channels) in the system to 1, and assume that all sub-channels are apportioned among all active and reachable MSs. For feasible group size $sz = (1,2,3)$ supported by the system, BS allots resources (i.e. sub-channels) according to ratio:

$$\lambda_1 : \lambda_2 : \lambda_3 = 1 : 3 : 7 = (2^1 - 1) : (2^2 - 1) : (2^3 - 1) \qquad (1)$$

We call the policy represented by Eq. (1) the Inter-coalition Binary Exponential Apportioning (Inter-BEA) mechanism. We comment that our approach is inherently efficient considering the above assumption, and we will further examine its effectiveness in subsection 3.1 by theorem 1 (in the sense of negative externality and no free-ride), in Section 5 (in the sense of coalition structure stability) and in Section 6 by simulation.

It's worth to point out that although we only present our research in terms of OFDMA in particular, Our incentive mechanism is in fact general enough to be applied in systems such as TDMA or CDMA, either by means of sub-slotting; i.e. the basic resource unit, the time-slot, is divided into source and relay subslots, or by means of sub-coding and multi-code transmissions.

Assume that MSs are all rational and selfish in that each of them would like to obtain more sub-channels and cost less energy. Given the apportioning policy of the BS, MSs in the outer ring would seek to find possible relays in middle or inner ring, so as to secure some communication opportunity; those in the inner ring would choose if they will relay for others and MSs in the middle ring can weigh the alternatives of whether or not to relay for MSs in outer ring, and whether or not to seek possible help from MSs in inner ring.

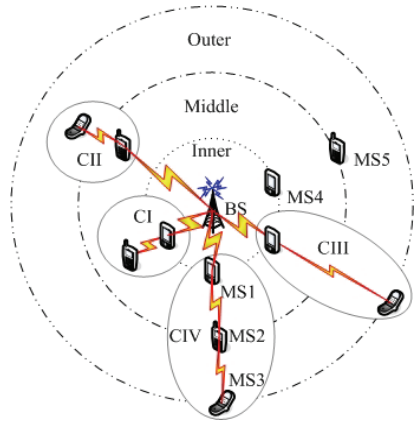

**Figure1**. An uplink multihop cellular network

## 3  Relaying cooperation game model & analyzing

### 3.1  Restricted relaying cooperation in partition function form

Under the particular configuration and Inter-group apportioning policy, rough examination shows that our model bears the following characteristics:

*Restricted cooperation* [12, 15, 16, 19, 23]. Since a source and its direct relay must within each other's transmission range, apparently a node can only cooperate with those falling within its neighbor area. Games with restricted cooperation describe situations in which the players are not completely free in forming coalitions; the restrictions in our model are mainly attributed to topology, namely the relative position of MSs. To manage these situations, a coalition set is generally introduced to represent the set of all feasible cooperative groups among the nodes.

*Externalities* [17, 20]: Noted that given the apportioning policy in Eq. (1), the action of merging (relaying coalition formation) among 2 or 3 MSs will affect payoffs of those nodes not involved in the mergence. These externalities can be captured in the framework of *partition function game (PFG)* [25] quite easily.

There can be two special cases of externalities: positive and negative externalities, meaning that when a mergence occurs, all the players who are not involved in a mergence are better (worse) off, more formally,

**Definition** 1(*Positive/negative externality games*[11,13,22,24]*)* the following condition characterizes a positive (negative) externality game: Assume that CS $\pi'$ was created from CS $\pi \neq \pi'$ by a mergence of certain (possibly also trivial) coalitions that consist of k players, $k \in \mathbf{N}$. If the *allocation function* $\phi(.)$ is defined in terms of payoffs: $\phi_i(\pi') \geq \phi_i(\pi)$ ($\phi_i(\pi') \leq \phi_i(\pi)$), $\forall i \in \mathbf{N} \setminus k$.

Combining the above judgment, we conclude that our model pertains to a kind of *PFG with restricted cooperation*, and we introduce a set of feasible *embedded coalitions* into a general PFG model to denote this restriction, this result in the following definition:

**Definition** 2(*RPFG—Relaying PFG)* A RPFG can be represented as $G(\mathbf{N},\mathbf{E},v,\phi)$, in which N denotes the set of MSs in MCN system shown in Fig.1, and $\mathbf{\Pi}(\mathbf{N}) = \{\{\mathbf{C}_1,...,\mathbf{C}_m\} | \mathbf{C}_i \cap \mathbf{C}_j = \mathbf{\Phi}, i \neq j; \cup_{i=1}^{m}\mathbf{C}_i = \mathbf{N}\}$ the set of feasible *coalition structures (CS)*. A pair E(C, π) with $\mathbf{C} \in \pi$ and $\pi \in \mathbf{\Pi}(\mathbf{N})$ is called an *embedded coalition (EC)* and the set of feasible embedded coalitions is denoted by $\mathbf{E}(\mathbf{N})$. The *partition function* **v** assigns a value to every feasible embedded coalition $v : \mathbf{E}(\mathbf{N}) \to \mathbf{R}$ according to apportioning policy in Eq. (1). An *allocation rule* is a function $\phi : (\mathbf{E},v) \to R^{\mathbf{N}}$ such that $\sum_i \phi_i(E,\mathbf{V}) = v(E)$ for all *v* and *E*.

Feasible coalitions in the RPFG model corresponding to the three type of feasible uplink transmitting mode mentioned in Sect. 2, namely single-hop, 2-hop and 3-hop transmission, hereafter also referred as singleton coalition (SC), double coalition (DC) and triple coalition (TC). To be noted that the value v(C, π) of the embedded coalition (C, π) is the payoff that the players in C can distribute among themselves in case the CS π forms. In other words, payoff is assumed to be transferable ([14] TU) within embedded coalitions, but nontransferable (NTU) between them. Through further examination, we reach the following conclusion about our RPFG model:

**Theorem** 1: For the non-trivial case of $n = |\mathbf{N}| > 2$, the RPFG model defined in definition 2 is a negative externality game as a whole.

*Proof of Theorem 1*: Suppose that before a mergence, the number of SC, DC and TC in the system are $n_1$, $n_2$ and $n_3$, $n_1 \geq 2, n_2 \geq 0, n_3 \geq 0$. According to apportioning policy (1), available sub-channels are divided into $m = \lambda_1 n_1 + \lambda_2 n_2 + \lambda_3 n_3 = n_1 + 3n_2 + 7n_3$, i.e., the allotting unit is $\sigma = 1/m$.

There can be all together 4 kinds of mergence: CI ~ CIV in Fig.1, in which CII ~ CIV involve an introduction of a MS within the outer ring (which is originally not an active node) while CI type not.

Firstly, if the mergence belongs to a CI type, before merging, the sub-channels allotted to the 2 involved nodes sum to $P = 2\sigma$. After the merging, the number of SC, DC and TC become $n_1' = n_1 - 2$, $n_2' = n_2 + 1$ and $n_3' = n_3$ respectively, $m' = \lambda_1 n_1' + \lambda_2 n_2' + \lambda_3 n_3' = n_1 + 3n_2 + 7n_3 + 1$, $\sigma' = 1/m'$, the sub-channels allotted to the newly formed DC (by the 2 merged nodes) are $P' = \lambda_2 \sigma' = 3\sigma'$. Solve $P' > P$, we get $n_1 + 3n_2 + 7n_3 > 2$., this can always be satisfied under the condition of $n = |\mathbf{N}| > 2$ ($n_1 > 2, n_2 = n_3 = 0$ or $n_1 = 2, n_2 \times n_3 > 0$), i.e., $P'$ is always greater than P, meaning that the mergence brings positive gain in payoffs for the involved members in a whole, since all available sub-channels are (normalized to)1, the payoffs to the residual nodes that not involved in the mergence apparently decrease after the merging.

Secondly, if the mergence belongs to CII or CIII type, before merging, the sub-channels allotted to the involved nodes are $P = \sigma$ (the node in outer ring is not reachable under SH and thus gets no sub-channels). After the merging, the number of SC, DC and TC become $n_1' = n_1 - 1$, $n_2' = n_2 + 1$ and $n_3' = n_3$, $m' = \lambda_1 n_1' + \lambda_2 n_2' + \lambda_3 n_3' = n_1 + 3n_2 + 7n_3 + 2$, $\sigma' = 1/m'$, the sub-channels allotted to the newly formed DC (by the 2 merged nodes) are $P' = \lambda_2 \sigma' = 3\sigma'$. Solve $P' > P$, we get $n_1 + 3n_2 + 7n_3 > 1$, this can always be satisfied under the condition of $n = |\mathbf{N}| > 2$, i.e., $P'$ is always greater than P, from now on the proof is similar to the CI case.

Lastly, if the mergence belongs to CIV type, before merging, the sub-channels allotted to the involved nodes are $P = 2\sigma$ (the node in outer ring gets no sub-channels while the one in inner and middle ring each gets $P = \sigma$). After the merging, the number of SC, DC and TC become $n_1' = n_1 - 2$, $n_2' = n_2$ and $n_3' = n_3 + 1$, $m' = \lambda_1 n_1' + \lambda_2 n_2' + \lambda_3 n_3' = n_1 + 3n_2 + 7n_3 + 5$, $\sigma' = 1/m'$, the sub-channels allotted to the newly formed TC (by the 3 merged nodes) are $P' = \lambda_3 \sigma' = 7\sigma'$. Solve $P' > P$, we get $n_1 + 3n_2 + 7n_3 > 1$ as in CII and CIII, from now on the proof is same as the CII and CIII case. This concludes the proof. ∎

It's well known in cooperative game theory domain that in a negative externality game there are no opportunities for nodes that don't cooperate to free-ride [11, 17]; this property provides motivation for nodes to form coalition.

### 3.2 Examples of three- and five-node cases

Consider the case when there are only three active nodes (i.e., |N|=3 and we only consider the special case of MS1, MS2 and MS3 as shown in fig. 1 here, each locates on the respective ring boundary) in the system, then the respective channel apportioning for all feasible ECs, i.e., the partition functions of game $G(\mathbf{N},\mathbf{E},v,\phi)$ are given as follows:

$$\mathbf{v}(1,2,3) = (1/2,1/2,0) \ ; \ \mathbf{v}(12,3) = (1,0) \ ; \ \mathbf{v}(1,23) = (1/4,3/4) \ ; \ \mathbf{v}(13,2) = (3/4,1/4) \ ;$$

$$\mathbf{v}(123) = (1) \tag{2}$$

In which zeros in $v(1,2,3)$ and $v(12,3)$ indicate that when no nodes relay for MS3, it can't reach BS by itself and gets no resources, so MS1 and MS2 have all the channels between them.

For the case when there are five nodes (i.e., |N|=5 and we only consider the special case of MS1~MS5 as shown in fig. 1 here, each locates on the respective ring boundary), the partition functions for all feasible CSs $\Pi = \{\pi = (\pi_1,\pi_2)\}$ are shown in Tab. 1, in which e.g., $\mathbf{v}(13,2,4,5) = (\frac{3}{6},\frac{1}{6},\frac{1}{6},\frac{1}{6})$.

Table 1. Partition function values for |N|=5 with Inter-BEA

| v(π) | | $\pi_1$ | | | | |
|---|---|---|---|---|---|---|
| | | 1,2,3 | 12,3 | 13,2 | 1,23 | 123 |
| $\pi_2$ | 4,5 | $\frac{1}{4},\frac{1}{4},0,\frac{1}{4},\frac{1}{4}$ | $\frac{3}{5},0,\frac{1}{5},\frac{1}{5}$ | $\frac{3}{6},\frac{1}{6},\frac{1}{6},\frac{1}{6}$ | $\frac{1}{6},\frac{3}{6},\frac{1}{6},\frac{1}{6}$ | $\frac{7}{9},\frac{1}{9},\frac{1}{9}$ |
| | 45 | $\frac{1}{5},\frac{1}{5},0,\frac{3}{5}$ | $\frac{1}{2},0,\frac{1}{2}$ | $\frac{3}{7},\frac{1}{7},\frac{3}{7}$ | $\frac{1}{7},\frac{3}{7},\frac{3}{7}$ | $\frac{7}{10},\frac{3}{10}$ |

## 4 What a fair Allocation should be - Prediction

Up to now, we have only proposed an allotting policy among coalitions for our MCN system, whether or not nodes will form relaying coalitions, and further which cooperative structure will be formed depends on how the worth v is divided among the nodes. Thus besides the inter-coalition policy, we also have to find an intra-coalition allocating mechanism. To get some insight in this, we would firstly try making some point prediction.

### 4.1 Prediction by Myerson value (MV) for PFG

The Shapley value (SV) [21] has been proven to be a useful solution concept for cooperative characteristic function TU games [25] (CFG, a special coalitional game without externalities) as it provides a recommendation for the division of the joint profits of the grand coalition, which assigns a unique allocation to each game, expresses compromise or average allocation and satisfies some reasonable properties.

The counterpart of SV for PFG was first proposed By R.B. Myerson [24], who showed that there exists the unique value which satisfies efficiency, symmetry and any one of the marginality axioms. The principle of marginality in PFG which states that one player's payoff should be calculated according to his marginal contribution. MV of PFG is defined as [24]:

$$\phi_i^{MV}(v) = \sum_{(C,\pi) \in E(N)} (-1)^{|\pi|-1}(|\pi|-1)!(\frac{1}{|N|} - \frac{1}{|\pi|-1} \sum_{\ni(\tilde{C} \neq C) \in p, i \notin \tilde{C}} \frac{1}{|N|-|\tilde{C}|}) v(C, \pi) \quad (3)$$

For N=3, the number of all possible CSs $|\Pi(N)|=5$, the number of all possible ECs $|E(N)|=10$. By Eq. (3) we get

$$\Phi^{MV}(v) = (11/24, 11/24, 1/12) = (0.458, 0.458, 0.083) \quad (4)$$

For N=5, the number of feasible CSs $|\Pi(N)|=10$, and the number of feasible ECs $|E(N)|=35$. Also by Eq. (3) we get

$$\Phi^{MV}(v) = (0.254, 0.254, 0.0808, 0.2056, 0.2056) \quad (5)$$

### 4.2 Prediction by Compensated Myerson Value (CMV)

Observe that under the MV mechanism in sub-section 4.1, payoff of MS1 in Eqs. (4) and (5) equals to that of MS2 and that of MS4 in Eq. (5) equals to MS5. We don't think this fair for our model as nodes in inner ring (e.g., MS1, MS4) generally cost more in relaying than those in middle ring. We in this sub-section design a way of compensation for forwarding cost on the basis of the MV prediction of sub-section 4.1.

Assume that each node uses equal *compensation weight* $\lambda \geq 0$ and $\Delta\phi_i = \lambda\phi_j$ means that node i gets $\lambda\phi_j$ additional Remuneration for its energy cost when it relays $\phi_j$ for node j. Introducing an indicator function as follows $I_{ij} = \begin{cases} 1, & i \text{ relays for } j \\ -1, & j \text{ relays for } i \\ 0, & \text{otherwise} \end{cases}$

and we define the MV after Compensation as:

$$\phi_i^{CC}(v) = \sum_{(\mathbf{C},\boldsymbol{\pi})\in E(\mathbf{N})} (-1)^{|\boldsymbol{\pi}|-1}(|\boldsymbol{\pi}|-1)!(\frac{1}{|\mathbf{N}|}-\frac{1}{|\boldsymbol{\pi}|-1}\sum_{\ni(\tilde{\mathbf{C}}\neq\mathbf{C})\in p, i\notin \tilde{\mathbf{C}}} \frac{1}{|\mathbf{N}|-|\tilde{\mathbf{C}}|})(v(\mathbf{C},\boldsymbol{\pi}) + \sum_{i,j\in\mathbf{C}} I_{ij}\phi_j^{CC}) \qquad (6)$$

By Eq. (6) we get the CMV with respect to $\lambda$ for $|N|=3$ as shown in Eq. (7) and Fig. 2.

$\phi_1^{CC} = \frac{1}{3}(v(123,\{123\}) + \lambda(\phi_2^{CC} + \phi_3^{CC}))$

$+\frac{1}{6}(v(12,\{12,3\}) + \lambda\phi_2^{CC}) - \frac{1}{3}v(3,\{12,3\})$

$+\frac{1}{6}(v(13,\{13,2\}) + \lambda\phi_3^{CC}) - \frac{1}{3}v(2,\{13,2\})$

$+\frac{2}{3}v(1,\{1,23\}) - \frac{1}{3}v(23,\{1,23\})$

$+\frac{1}{6}v(2,\{1,2,3\}) + \frac{1}{6}v(3,\{1,2,3\}) - \frac{1}{3}v(1,\{1,2,3\})$

$\phi_2^{CC} = \frac{1}{3}(v(123,\{123\}) + \lambda(-\phi_2^{CC} + \phi_3^{CC}))$

$+\frac{1}{6}(v(12,\{12,3\}) - \lambda\phi_2^{CC}) - \frac{1}{3}v(3,\{12,3\})$

$+\frac{1}{6}(v(23,\{1,23\}) + \lambda\phi_3^{CC}) - \frac{1}{3}v(1,\{1,23\})$

$+\frac{2}{3}v(2,\{13,2\}) - \frac{1}{3}v(13,\{13,2\})$

$+\frac{1}{6}v(1,\{1,2,3\}) + \frac{1}{6}v(3,\{1,2,3\}) - \frac{1}{3}v(2,\{1,2,3\})$

$\phi_3^{CC} = \frac{1}{3}(v(123,\{123\}) - 2\lambda\phi_3^{CC})$

$+\frac{1}{6}(v(13,\{13,2\}) - \lambda\phi_3^{CC}) - \frac{1}{3}v(2,\{13,2\})$

$+\frac{1}{6}(v(23,\{1,23\}) - \lambda\phi_3^{CC}) - \frac{1}{3}v(1,\{1,23\})$

$+\frac{2}{3}v(3,\{12,3\}) - \frac{1}{3}v(12,\{12,3\})$

$+\frac{1}{6}v(1,\{1,2,3\}) + \frac{1}{6}v(2,\{1,2,3\}) - \frac{1}{3}v(3,\{1,2,3\})$

, $\begin{cases} \phi_3 = \frac{1}{12(1+\lambda)} \\ \phi_2 = \frac{11+12\lambda}{12(1+\lambda)(2+\lambda)} \\ \phi_1 = 1 - \varphi_2 - \varphi_3 \end{cases}$ (7)

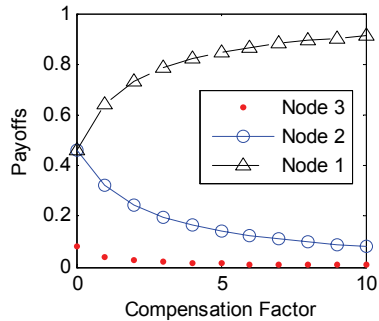

**Figure 2**. Payoff of each MS predicted by CMV vs. *compensation weight* $\lambda$ for $|N|=3$

We see from Fig.2 that there in fact can be many ways of tradeoff in payoff allocation when compensation for relaying cost is taken into consideration.

## 4.3 Proposed intra-coalition compensation scheme: Intra-BEA

Though Myerson value given in Sect.4 can provide some insight in fair payoff allocation, yet in a practical system where there are many nodes, MV-like payoff allocating methods are not flexible, owing to the complexity and rigid Inter-dependence among all nodes. In fact it's even not applicable in our system, as on the one hand, it's difficult for a MS to know exactly the global apportioning so as to calculate the fair MV, on the other hand, payoffs are not transferable among coalitions inside a particular CS. To overcome this drawback, we propose the following distributed intra-coalition allocation method which only requires local knowledge.

In a multihop uplink coalition, let the source, relay-I and relay-II (e.g. MS3, MS2 and MS1 in CIV in fig.1) share channels in ratio

$$\hbar_3 : \hbar_2 : \hbar_1 = 1 : 2 : 4 = 2^0 : 2^1 : 2^2 \tag{8}$$

We refer Eq. (8) as the Rule of Intra-coalition Binary Exponential Allocation (Intra-BEA). Given the system configuration and global policy Inter-BEA in Section 2, this local rule can be physically understood as follows: without loss of generality suppose $\lambda_1 = 1$, then according to Inter-BEA Eq. (1), a CIV-like coalition whose size is 3 gets 7 sub-channels totally, then MS1 can transmit to BS on all of them, in which 4 of them are traffics of its own, 2 are traffics of MS2, and 1 of them is MS3's. A DC type coalition (such as CI, CII and CIII in fig. 1 whose size is 2) gets 3 sub-channels, then traffics of the helping node and the helped one occupy 2 and 1 respectively. One interpretation of this Intra-BEA rule is that '*You reap what you sow*', i.e., each active node originally has one share of sub-channels, relay-I gets one extra reward (sub-channel) for forwarding on one sub-channel for the source and relay-II acquires 3 extra rewards for forwarding on 3 sub-channels (one for the source and the other two for relay-I), hence we get Eq. (8). Using Inter- and Intra-BEA, i.e., referring to Eq. (2), Tab.1 and Eq. (8) to the case of |N|=3 and |N|=5, we have the payoff of each node (per-membership allocation) for |N|=3 in Tab.2 and |N|=5 in Tab.4. 'Total-traffic' in Tab.3 and 5 means all the sub-channels a node need to transmit on, including self-traffics and those it relays for others.

Table 2. Payoff/self-traffic $\phi_i(\pi)$ of each node vs. CS for |N|=3 with Inter- & Intra-BEA

| Payoff | MS1 | MS2 | MS3 |
|---|---|---|---|

|   | 1,2,3 | 1/2 | 1/2 | 0 |
|---|---|---|---|---|
|   | 12,3 | 2/3 | 1/3 | 0 |
|   | 13,2 | 1/2 | 1/4 | 1/4 |
|   | 1,23 | 1/4 | 1/2 | 1/4 |
|   | 123 | 4/7 | 2/7 | 1/7 |

**Table 3**. Total-traffic $t_i(\pi)$ of each node vs. CS for $|N|=3$ with Inter- & Intra-BEA

| Total | MS1 | MS2 | MS3 |
|---|---|---|---|
| 1,2,3 | 1/2 | 1/2 | 0 |
| 12,3 | 1 | 1/3 | 0 |
| 13,2 | 3/4 | 1/4 | 1/4 |
| 1,23 | 1/4 | 3/4 | 1/4 |
| 123 | 1 | 3/7 | 1/7 |

**Table 4**. Payoff/self-traffic $\phi_i(\pi)$ of each node vs. CS for $|N|=5$ with Inter- & Intra-BEA

| Payoff | | $\pi_1$ | | | | |
|---|---|---|---|---|---|---|
| | | 1,2,3 | 12,3 | 13,2 | 1,23 | 123 |
| $\pi_2$ | 4,5 | $\frac{1}{4},\frac{1}{4},0,\frac{1}{4},\frac{1}{4}$ | $\frac{2}{5},\frac{1}{5},0,\frac{1}{5},\frac{1}{5}$ | $\frac{2}{6},\frac{1}{6},\frac{1}{6},\frac{1}{6},\frac{1}{6}$ | $\frac{1}{6},\frac{2}{6},\frac{1}{6},\frac{1}{6},\frac{1}{6}$ | $\frac{4}{9},\frac{2}{9},\frac{1}{9},\frac{1}{9},\frac{1}{9}$ |
| | 45 | $\frac{1}{5},\frac{1}{5},0,\frac{2}{5},\frac{1}{5}$ | $\frac{1}{3},\frac{1}{6},0,\frac{1}{3},\frac{1}{6}$ | $\frac{2}{7},\frac{1}{7},\frac{1}{7},\frac{2}{7},\frac{1}{7}$ | $\frac{1}{7},\frac{2}{7},\frac{1}{7},\frac{2}{7},\frac{1}{7}$ | $\frac{4}{10},\frac{2}{10},\frac{1}{10},\frac{2}{10},\frac{1}{10}$ |

**Table 5**. Total-traffic $t_i(\pi)$ of each node vs. CS for $|N|=5$ with Inter- & Intra-BEA

| Total | | $\pi_1$ | | | | |
|---|---|---|---|---|---|---|
| | | 1,2,3 | 12,3 | 13,2 | 1,23 | 123 |
| $\pi_2$ | 4,5 | $\frac{1}{4},\frac{1}{4},0,\frac{1}{4},\frac{1}{4}$ | $\frac{3}{5},\frac{1}{5},0,\frac{1}{5},\frac{1}{5}$ | $\frac{3}{6},\frac{1}{6},\frac{1}{6},\frac{1}{6},\frac{1}{6}$ | $\frac{1}{6},\frac{3}{6},\frac{1}{6},\frac{1}{6},\frac{1}{6}$ | $\frac{7}{9},\frac{3}{9},\frac{1}{9},\frac{1}{9},\frac{1}{9}$ |
| | 45 | $\frac{1}{5},\frac{1}{5},0,\frac{3}{5},\frac{1}{5}$ | $\frac{1}{2},\frac{1}{6},0,\frac{1}{2},\frac{1}{6}$ | $\frac{3}{7},\frac{1}{7},\frac{1}{7},\frac{3}{7},\frac{1}{7}$ | $\frac{1}{7},\frac{3}{7},\frac{1}{7},\frac{3}{7},\frac{1}{7}$ | $\frac{7}{10},\frac{3}{10},\frac{1}{10},\frac{3}{10},\frac{1}{10}$ |

## 5 CS Stability with respect to the Inter- & Intra-BEA allocation mechanism

Given apportioning and compensating schemes in Sect. 2 and 4, we are interested in which CS is stable, if any, i.e., what coalitions the nodes would like to form. The stability of a CS is justified in terms of Internal and external stability [13] for $|N|=3$ and inductive core [18, 22] for $|N|=5$.

## 5.1 Utilities of nodes given Inter- & Intra-BEA

It should be pointed out that payoffs in form of sub-channels allocated alone are not sufficient to represent the gain and loss of a MS in the game, as the introduction of multihop affects the transmitting power $p_i(\pi)$, and further the energy cost $c_i(\pi)$ of each involved MS i, which equals to the product of $p_i(\pi)$ and $t_i(\pi)$, i.e., $c_i(\pi) = p_i(\pi)t_i(\pi)$.

To be noted again that MS1~MS5 all locate on respective ring boundary. Assume that channel gain of a MS i is modeled by $h_i = r_i^{-a}$ [27], in which $r_i$ is the distance between MS i and BS. Then the respective channel power gain $|h_i|^2 = r_i^{-2a}$. Further assume that each ring ($R_I$, $R_M$ and $R_O$) in fig. 1 has equal width, i.e., $w_I = w_M = w_O = \frac{1}{3}r_{BS}$ and $a = 2$. Assume perfect power control, BS and all MSs use same modulation schemes, if we normalize the transmitting power (needed in one sub-channel for successful receiving) of a node on the outer boundary of $R_M$ (e.g., MS2) when transmitting directly to BS to 1, then the necessary transmitting power of MS2 in one sub-channel when communicating to MS1 is only $2^{-4}$. We get transmitting power of each node in all feasible CSs for |N|=3 as shown in tab. 6, and for |N|=5 in tab.7.

**Table 6**. Transmitting power $p_i(\pi)$ of each node vs. CS for |N|=3

| Power | MS1 | MS2 | MS3 |
|---|---|---|---|
| 1,2,3 | $2^{-4}$ | 1 | 0 |
| 12,3 | $2^{-4}$ | $2^{-4}$ | 0 |
| 13,2 | $2^{-4}$ | 1 | 1 |
| 1,23 | $2^{-4}$ | 1 | $2^{-4}$ |
| 123 | $2^{-4}$ | $2^{-4}$ | $2^{-4}$ |

**Table 7**. Transmitting power $p_i(\pi)$ of each node vs. CS for |N|=5

| Power | | $\pi_1$ | | | | |
|---|---|---|---|---|---|---|
| | | 1,2,3 | 12,3 | 13,2 | 1,23 | 123 |
| $\pi_2$ | 4,5 | $2^{-4},1,0,2^{-4},1$ | $2^{-4},2^{-4},0,2^{-4},1$ | $2^{-4},1,1,2^{-4},1$ | $2^{-4},1,2^{-4},2^{-4},1$ | $2^{-4},2^{-4},2^{-4},2^{-4},1$ |
| | 45 | $2^{-4},1,0,2^{-4},2^{-4}$ | $2^{-4},2^{-4},0,2^{-4},2^{-4}$ | $2^{-4},1,1,2^{-4},2^{-4}$ | $2^{-4},1,2^{-4},2^{-4},2^{-4}$ | $2^{-4},2^{-4},2^{-4},2^{-4},2^{-4}$ |

Apparently, each node cares not only gains in payoff but also loss in energy. They would like to get more sub-channels (payoffs) with less energy cost. Accordingly, we define the utility $U_i(\boldsymbol{\pi})$ of MS i as the weighted difference between its payoff (sub-channels allocated) $v_i(\boldsymbol{\pi})$ and its energy cost $c_i(\boldsymbol{\pi})$:

$$U_i(\boldsymbol{\pi}) = \rho \phi_i(\boldsymbol{\pi}) - (1-\rho)c_i(\boldsymbol{\pi}), 0 < \rho < 1 \tag{9}$$

From tab. 2, tab.3 and tab.6, by Eq. (9), we find utility of each node for |N|=3 as shown in tab. 7, and from tab.4, tab.5 and tab.8 we find utility in five-node case as shown in tab. 9. Note that there appear negative utilities in both table, e.g., $U_2(1,23) = -0.125$, $U_2(1,23,4,5) = -0.083$ and $U_2(1,23,4,5) = -0.071$. These should indicate that in these cases, the energy cost of MS2 is so large that its gains in payoffs are not enough to countervail it.

**Table 8**. Utility $U_i(\boldsymbol{\pi})$ of each node vs. CS for |N|=3 with Inter- & Intra-BEA

| Utility | MS1 | MS2 | MS3 |
|---|---|---|---|
| 1,2,3 | 0.234 | 0 | 0 |
| 12,3 | 0.302 | 0.156 | 0 |
| 13,2 | 0.227 | 0 | 0 |
| 1,23 | 0.117 | -0.125 | 0.117 |
| 123 | 0.254 | 0.129 | 0.067 |

**Table 9**. Utility $U_i(\boldsymbol{\pi})$ of each node vs. CS for |N|=5 with Inter- & Intra-BEA

| Utility | | $\pi_1$ | | | | |
|---|---|---|---|---|---|---|
| | | 1,2,3 | 12,3 | 13,2 | 1,23 | 123 |
| $\pi_2$ | 4,5 | 0.117, 0, 0 | 0.181, 0.094, 0 | 0.151, 0, 0 | .078, -.083, .078 | 0.198, 0.101, 0.052 |
| | | 0.117, 0 | 0.094, 0 | 0.078, 0 | 0.078, 0 | 0.052, 0 |
| | 45 | 0.094, 0, 0 | 0.151, 0.078, 0 | 0.129, 0, 0 | .067, -.071, .067 | .178, .091, .047 |
| | | 0.181, 0.094 | 0.151, 0.078 | 0.129, 0 | 0.129, 0.067 | 0.091, 0.047 |

## 5.2 CS stability for |N|=3

It's apparent that all CSs for |N|=3 are single-agreement CS (in which there is only one (non-trivial) coalition), so we can investigate its stability based on the conception of internal and external stability [13]. A CS is said to be stable if it is both internally and externally stable.

***Definition*** 3 (*Internal* and *external stability*) A CS $\boldsymbol{\pi} := [\mathbf{C} | \text{singletons}]$ is:

(a) Internally stable if and only if $U_i([\mathbf{C}|\text{singletons}]) \geq U_i([\mathbf{C}_{-i}|\text{singletons}|i])$, $\forall i \in \mathbf{C}$; and

(b) Externally stable if and only if $U_i([\mathbf{C}|\text{singletons}]) < U_i([\mathbf{C} \cup \{i\}|\text{singletons}_{-i}])$, $\forall i \in \text{singletons}$.

Firstly referring to Tab.8 we see that while [1|23] is the best choice of MS3, it is the last choice of MS2, and so it's not internally stable. MS2 would break from [1|23] to form [1|2|3] and further to cooperate with MS1 to form [12|3], which is the best choice for both MS1 and MS2, indicating that this CS is internally stable. CS [13|2] is not internally stable similarly as [1|23]. It's noticeable that though the grand coalition [123] isn't internally stable, it's the second best choice for all the members.

## 5.3 CS stability for |N|=5

For |N|=5, as there can be more than one non-trivial coalition in a particular CS, the model belongs to multiple-agreement games and stability conceptions like *Internal* and *external stability* aren't applicable here[13]. We adopt instead the inductive Core [18, 22] to investigate stability of multiple-agreement CS in this case.

Informally, the core [22, 23] is the set of un-dominated allocations. An allocation is not in the core if there is a coalition that can profitably deviate from it. In PFGs a deviation by a coalition or a set of coalitions typically affects the payoffs of the residual players thus invoking a response from them. Such a response can change the worth of the deviation dramatically. L, Kóczy in [18, 22] defines a Core C+ by induction based on a dominance in which residuals form an allocation that is a member of the core in the residual subgame in a reaction to a deviation. Formally, the definition consists of four steps:

***Definition* 4 (*the inductive Core C+*)**

1. *Trivial game*. Let $G(\mathbf{N},\mathbf{E},v,\phi)$ be a PFG game. The core of a game with N = {1} is the only allocation with the trivial CS: $\mathbf{C}_+(\{1\},v) = \{(v(1,(1)),(1))\}$.

2. *Inductive assumption*. Given the definition of the Core $\mathbf{C}_+(\mathbf{R},v)$, in which $\mathbf{R} = \mathbf{N} \setminus \mathbf{C}$ are the set of residual nodes for every game with at most k + 1 nodes, we can define dominance for a game of k players. Let $\mathbf{A}_+(\mathbf{R},v)$ denote the optimistic assumption about the game $G(\mathbf{R},\mathbf{E_R},v,\phi_\mathbf{R})$. If $\mathbf{C}_+(\mathbf{R},v) \neq \Phi$ then $\mathbf{A}_+(\mathbf{R},v) = \mathbf{C}_+(\mathbf{R},v)$ otherwise $\mathbf{A}_+(\mathbf{R},v) = \Omega_+(\mathbf{R},v)$, i.e., all the feasible allocation vectors.

3. *Dominance*. The allocation $(x,\pi)$ is dominated via the coalition C forming CS $\pi$ if there exists an assumption $(\phi_{\mathbf{N}\setminus\mathbf{C}},\pi_\mathbf{R}) \in \mathbf{A}_+(\mathbf{N}\setminus\mathbf{C},v_\mathbf{C})$ and an allocation $((\phi_C, \phi_{N \setminus C}), \pi_C \cup \pi_R)$ such that $\phi >_C x$. The allocation $(x, \pi)$ is dominated if it is dominated via a coalition.

4. *Core*. The core of a game of k players is the set of un-dominated allocations and we denote it by $C_+(N, v)$.

Referring to Tab.9, same as three-node case, we see that while CS $[1|23|4|5]$ is the best choice of MS3, it's the last choice of MS2; the case of CS $[1|23|4|5]$ is similar. Note that CS $[123|4|5]$ pertains to single-agreement CS and is internally stable for all members in EC $(123, [123|4|5])$, which is the best choice of MS1, MS2 and MS3. CS $[1|2|3|45]$ is just a counterpart of $[123|4|5]$ in that it's the best choice of MS4 and MS5. From Tab.9 it's apparent that when MS1, MS2 and MS3 choose to form coalition {123}, the best response of the residual nodes (i.e., MS4 and MS5) is to form coalition {45}, which results in CS $[123|45]$, namely, the inductive Core of the five-node PFG.

To conclude, for |N|=3, CS $[12|3]$ is internally stable; CS $[123]$ is the second best choice for all its members. For |N|=5, CS $[123|45]$ is the inductive Core as well as the second best choice for all the nodes. This shows that our BEA incentive measure is indeed effective in these special cases.

## 6  Simulation and numerical results

In the former sections we focus on special cases such as nodes all located on the boundaries (e.g., MS1~MS5 in Fig.1). In this section, we examine the applicability and efficiency (in terms of network connectivity and nodes' payoff and energy cost etc.) of the proposed BEA stimulating approach in more general cases (more nodes, random location) by means of simulation. The problems that we try to answer in this section are, firstly from the point of the system, on what scale can the proposed scheme improve connectivity? Secondly from the view of individual node, what is the gain and loss to cooperate in terms of payoff and energy cost?

### 6.1  Simulation configuration

Consider a cell with BS centered at the origin and MSs uniformly distributed in the area. The number of MSs in the inner, middle and outer ring (abbreviated $R_I$, $R_M$, $R_O$ respectively) is denoted as $\mathbf{M} = (M_j, j = 1 \sim 3)$ and the number of SC, DC and TC is denoted as $\mathbf{N} = (N_i, i = 1 \sim 3)$. According to allotting policy (1), we have $N_1 + 2N_2 + 3N_3 = \sum_{j=1}^{3} M_j$ and $\sum_{i=1}^{3} \lambda_i N_i = 1$. We use the same channel model as in sub-section 5.1 and some default simulation parameters are given in Tab.10.

Table 10. Simulation parameters

| Parameters | Value |
|---|---|
| Cell diameter (m) | D=500 |
| Ring width (m) | D/6 |
| Service Area diameter (m) | 1.1D/3 |
| Variance of Gaussian noise $N_0$ (dbW) | -133 |
| Path loss exponent α | 2 |
| SNR threshold of successful decoding (dB) | 10 |
| Number of MSs | 60 |
| Realizations of Monte Carlo simulation | 50 |

## 6.2 Coalition formation algorithm

As coalitions of CIII type are always dominated by those of CIV type (referring to $CS_{[13|2]}$ vs. $CS_{[123]}$ in Tab.8 and $CS_{[13|2|...]}$ vs. $CS_{[123|...]}$ in Tab.9), we omit this kind of coalitions in the coalition formation. Given Inter- and Intra-BEA in sub-section 2.2 and 4.3, assume that nodes in inner and middle ring are always willing to forward (for those applicants fall within their *service area,* which are disks centered at the respective relays) and they form coalition according to rules as follows:

As source, a node in outer or middle ring broadcasts its request, and then chooses from the respondents the one with best channel.

As relay, a node in inner or middle ring will accept a request if the applicant located within its service area and it has not yet form a coalition with any source.

## 6.3 Monte Carlo simulation

We define *network connectivity* as the ratio of the number of nodes which can reach BS to the total number of nodes in the system, the *efficiency* of a node as the ratio of its payoff to its energy cost, and the *MCN relative efficiency* of a node as the ratio of its efficiency in MCN to its efficiency in a single-hop network. We investigate the above performance metrics in terms of various nodes' density (i.e., number of nodes), size of service area and ring width. 'Width of service area' which denoted as (2, 21/10, 11/5, 12/5, 3) in fig. 3(b) and 6, 7 are the ratio to ring width (500/6m), so the real values are (166. 7, 175.0, 183.3, 200. 0, 250.0) respectively. 'I' and 'M' in Fig.4~9 denotes nodes in inner and middle ring respectively, 'M-CIV' denotes nodes locate in middle ring who take part in coalitions of CIV type.

Table 11. *Difference in ring width* in fig. 3(c) and 8, 9 vs. width of inner, middle and outer ring

| difference | inner | middle | outer |
|---|---|---|---|
| -1/16 | 114.5833 | 83.3333 | 52.0833 |
| -1/32 | 98.9583 | 83.3333 | 67.7083 |
| 0 | 83.3333 | 83.3333 | 83.3333 |
| 1/32 | 67.7083 | 83.3333 | 98.9583 |
| 1/16 | 52.0833 | 83.3333 | 114.5833 |

**Network connectivity**

In fig.3, 'gain of coop' denotes the connectivity difference between cooperative and non-cooperative models. It can be seen that our scheme can generally provides about 20% improvement, and it works better when the number of nodes increase, or when relays set service area bigger, or when the inner ring is wider than outer ring. Note that improvement in connectivity also represents how many nodes in outer ring benefit form forwarding cooperation and thus can set up connection with BS. It might be surmised that if we allow more coalition form (e.g., a relay can forward for multiple nodes in outer ring), the connectivity can be still better.

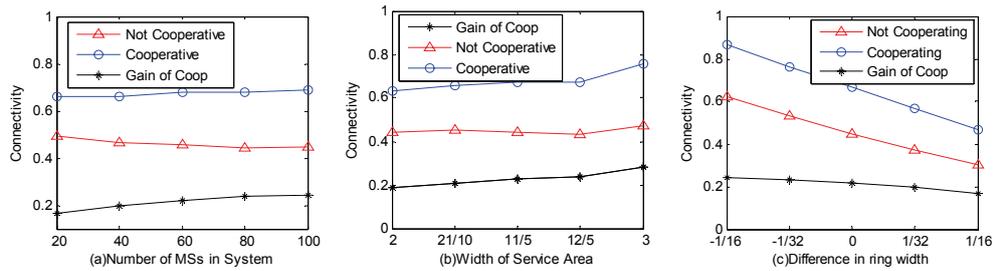

**Figure 3**. Network connectivity

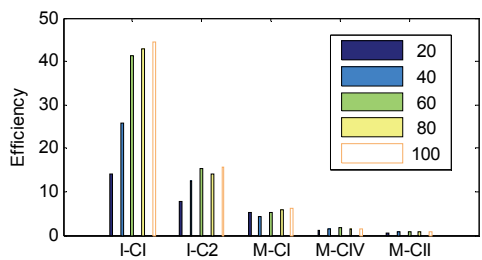

**Figure 4**. Efficiency vs. number of nodes

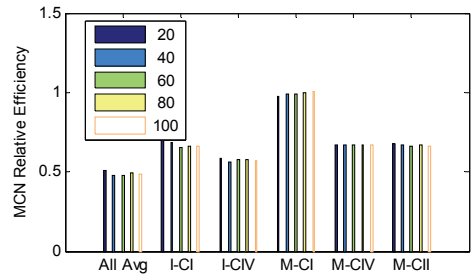

**Figure 5**. Relative Efficiency vs. number of nodes

*Efficiency* of nodes in inner or middle ring

We see from Fig.4, 6 and 8 that inner nodes generally have a better efficiency with respect to ratio of payoff to energy cost, and the efficiency increases with nodes density and service area size, or when outer ring is wider than inner ring.

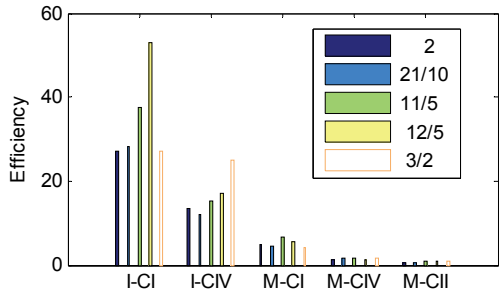

**Figure 6**. Efficiency vs. service area diameter

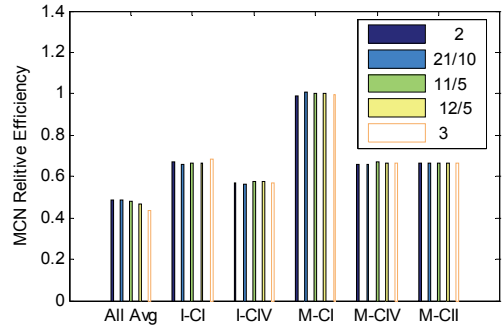

**Figure 7**. Relative Efficiency vs. service area diameter

*Relative efficiency* of nodes in inner or middle ring

'All Avg' in Fig. 5, 7 and 9 denotes the ratio of the average efficiency of all active nodes (suppose the number is X) in MCN mode to that of all active nodes (suppose the number is Y) in single-hop mode, note that X is greater than Y and the difference between them is the number of outer ring nodes that are helped in MCN mode. Note that middle inner nodes have the highest gain for participating CI type coalitions which should be attributed to the saving in energy for being relayed. Note also that nodes in inner and middle ring participating in whatever coalitions have a better relative efficiency than that of 'All Avg', which includes nodes within outer ring and this shows that nodes in outer ring have lowest efficiency. The effect of nodes density, service area size and ring width on relative efficiency is generally neglectable.

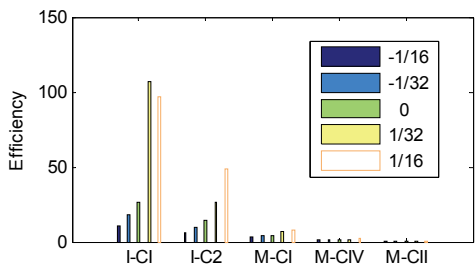

**Figure 8**. Efficiency vs. ring width

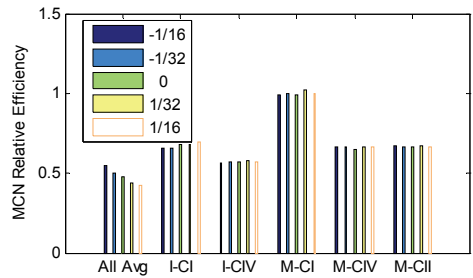

**Figure 9**. Relative Efficiency vs. ring width

# 7 Conclusion

Different from existent researches, we propose an alternative forwarding incentive approach dispensed with indirect monetary rewards, which is simple and light weighted in terms of implementation and maintenance, thus possibly sheds light on stimulating measure design in temporary or emergent circumstances with frequent entry and exit. Analysis and simulation results show that our measure can achieve some tradeoff between connectivity of nodes in outer ring (i.e., network coverage) and utilities of potential relays (i.e., nodes in inner and outer ring).